\documentclass[letterpaper,twocolumn,10pt]{article}
\usepackage{usenix2021_SOUPS}

\usepackage{tikz}
\usepackage{amsmath}

\usepackage{breakurl}

\begin{document}

\date{}

\title{\Large \bf Examining the Landscape of Digital Safety and Privacy Assistance \\ for Black Communities}

\def\plainauthor{Nikita Samarin, Aparna Krishnan, Moses Namara, Joanne Ma, Elissa M. Redmiles}

\author{
{\rm Nikita Samarin}\\
UC Berkeley
\and
{\rm Aparna Krishnan}\\
UT Austin
\and
{\rm Moses Namara}\\
Clemson University
\and
{\rm Joanne Ma}\\
UC Berkeley
\and
{\rm Elissa M. Redmiles}\\
Max Planck Institute for Software Systems
} 

\maketitle
\thecopyright

\begin{abstract}
Recent events have placed a renewed focus on the issue of racial justice in the United States and other countries. One dimension of this issue that has received considerable attention is the security and privacy threats and vulnerabilities faced by the communities of color. 

Our study focuses on community-level advocates who organize workshops, clinics, and other initiatives that inform Black communities about existing digital safety and privacy threats and ways to mitigate against them. Additionally, we aim to understand the online security and privacy needs and attitudes of participants who partake in these initiatives. We hope that by understanding how advocates work in different contexts and what teaching methods are effective, we can help other digital safety experts and activists become advocates within their communities.
\end{abstract}

\section{Introduction}

After the death of George Floyd in May 2020, a series of large-scale protests against police brutality and racism took place across the cities in the United States and later in other countries~\cite{lee2020demographics}. The demonstrations, as well as the law enforcement response against the protesters, have underscored the security and privacy threats faced by communities of color~\cite{steward2020systemic}. 

However, these threats are anything but new. Over the years, researchers in various fields have published numerous articles, reports, and journalistic pieces discussing the disproportionate impact of digital safety and privacy threats on the communities of color, including the heightened risk of surveillance of online~\cite{patton2017stop} and real-world activities~\cite{shere2020police}, online harassment~\cite{rector2020caller}, propaganda and disinformation~\cite{roose2020black}, and biases in algorithmic decision-making~\cite{grother2019face}. These threats, which heighten existing inequalities and often result from actions by law enforcement and other state actors~\cite{weitzer2006race}, violate the principles laid out in the United Nations Universal Declaration of Human Rights~\cite{assembly1948universal}.

This study aims to understand the existing approaches to educating members of Black communities in the United States about their online security and privacy and assess how effective these approaches are. As a case example, we focus on community-level advocates who organize workshops, clinics, and other initiatives that inform members of Black communities about existing digital safety and privacy threats and ways to mitigate against them. Through interviews with advocates from civil society groups and grassroots movements that focus on the issues of digital safety and racial injustice, we plan to answer the following research questions:

\begin{itemize}
  \item \textbf{RQ1}: What are the practices that advocates use to organize digital safety and privacy initiatives for members of Black communities?
  \item \textbf{RQ2}: What are the challenges that advocates encounter when organizing these initiatives?
  \item \textbf{RQ3}: How do advocates measure the success of the initiatives that they organize?
  \item \textbf{RQ4}: What is the efficacy of these initiatives in improving the digital safety and privacy for their intended audience, as perceived by the advocates and the participants of these initiatives?
\end{itemize}

Additionally, we aim to observe these practices \textit{in-situ} by attending events hosted by the advocates and understand the online security and privacy needs and attitudes of participants who take part in these events. We hope that by understanding how advocates work in different contexts and what teaching methods are effective, we can help other digital safety experts and activists become advocates within their communities.

\section{Related Work}
Prior work has demonstrated variability in online security and privacy experiences and concerns based on race and ethnicity. A 2019 study by the Pew Research Center has reported that Black Americans are more likely to experience social media or email breaches than their white counterparts (20\% vs. 6\%)~\cite{auxier2019americans}. There are also differences in the perceived threat actor of security and privacy risks, with Black Americans being more likely than white Americans to believe that the government is tracking their activities online (60\% vs. 43\%)~\cite{auxier2019americans} and more than three times as likely to say they are concerned about being unfairly targeted by law enforcement (73\% vs. 23\%)~\cite{madden2017privacy}. Finally, there is a difference in the impact of security and privacy threats on members of different racial and ethnic groups. For instance, Black Americans are more likely to face discrimination by algorithmic decision-making processes, leading to a reduction in healthcare~\cite{obermeyer2019dissecting} or a higher risk of being misidentified by a facial recognition algorithm in a criminal investigation case~\cite{klare2012face}. 

We believe that the variability of these factors and the unique nature of digital safety threats translate into differences between the security and privacy assistance and education required for racial and ethnic minority groups compared to other populations. Furthermore, a recent study by Boyd et al. examining the security and privacy advice given to Black Lives Matter protesters has emphasized the need for future work to “further investigate and codify best practices for [...] community-based activist trainings”~\cite{boyd2021understanding}.


Some studies have also explored best practices for improving the digital safety of targeted individuals and groups by providing personalized assistance from trained teams of technologists and security experts. Such computer security clinics currently exist for victims of intimate partner violence~\cite{havron2019clinical}, journalists~\cite{committee2020cpj}, activists~\cite{scott2016security}, and politically-vulnerable organizations~\cite{brooks2018defending}. Despite the increase in the prevalence of such research, no academic study known to us has explored the best practices for improving the security and privacy of Black communities in the United States using a participatory research design. Our work, therefore, will contribute to the academic literature on the security and privacy needs of underserved communities and serve as a step towards the reduction of racial inequality caused by the disproportionate impact of security and privacy threats.

\section{Methodology}
The study consists of three parts: an observational study, interviews with advocates, and interviews with participants. We will complete all parts of the study remotely to mitigate COVID-19 transmission risks.

To achieve our research goals, we will be closely collaborating with Matt Mitchell and Sarah Aoun from CryptoHarlem, a digital surveillance clinic that organizes impromptu workshops teaching basic cryptography tools to the predominately African American community in upper Manhattan.\footnote{\url{https://www.cryptoharlem.com/}} Mitchell and Aoun will discuss their experiences leading workshops at CryptoHarlem and, through snowball sampling, introduce us to other community-level advocates and facilitate the recruitment of participants for our interviews.

\subsection{Observational Study}
Alongside the interviews, our research team will remotely participate in the workshops, clinics, and other initiatives hosted by advocates.~\footnote{These initiatives have been traditionally held in person but have since moved online due to the ongoing COVID-19 pandemic, providing the opportunity for us to attend these events remotely.} We shall adopt a "fly-on-the-wall" observation technique in order to note any relevant practices that occur during those events, such as the topic of the session, online platform used for content delivery, duration of the event, type of interactions between the audience and the facilitator, and so on. The findings from this part of the study will complement the results that we obtain from the interviews. 

To ensure that we only perform observation of public behavior, we will observe events that are accessible to anyone who has a computer, Internet access, and, possibly, a social media account. Examples of such events include educational videos live-streamed on Twitch, Youtube, or Facebook available to all platform members. We will use the same criteria when watching the recordings of past events.

\subsection{Interviews with Advocates}
To gain a better understanding of the normative landscape of approaches used to inform members of Black communities about online security and privacy threats, we will perform an interview study with digital-safety advocates working at the community level. These interviews will also allow us to learn about the challenges that advocates face when they organize educational activities and the metrics that they use to measure the success of their efforts.

\subsubsection{Subject Population}

We will use the following criteria to select advocates to interview:

\begin{enumerate}
\item An “advocate” organizes, helps organize, or works with someone directly who organizes initiatives, including but not limited to: workshops, meetups, trainings, conferences, clinics, consultations, and other types of events and activities.
\item The organized initiative focuses at least partially on digital safety, online security, or privacy threat mitigation strategies applicable for individuals and communities. 
\item The advocates will decide themselves whether their initiative focuses on digital safety and privacy, to avoid the researchers imposing a definition onto them. 
\item The target audience for the initiative should primarily include members of a Black community.
\item The initiative takes place on a regular basis, i.e., is held more than once with different participants. 
\item The initiative takes place within the borders of the U.S. (if in person), or the advocate resides in the U.S. (if online).
\end{enumerate}

We expect to interview at least 12 advocates, who will be initially identified by our partner organization CryptoHarlem. Mitchell and Aoun, who lead CryptoHarlem, will send prospective participants an invitation to complete a screening survey to determine their eligibility according to the criteria that we outline above. Respondents will also answer demographic questions and indicate their preferred contact information, allowing us to reach out to them directly to schedule the interview. We will compensate each advocate with \$30 once we complete the interview and ask them to invite other prospective participants.

\subsubsection{Interview Guide}
We will conduct semi-structured interviews lasting around 50 minutes to gain a better understanding of the normative landscape of approaches used to inform members of Black communities about online security and privacy threats. We will focus on four topics relevant to our research questions:

\begin{itemize}
\item \textbf{Motivation.}
\begin{itemize}
    \item How did you first get involved with this type of work?
    \item What does ‘digital safety and privacy’ mean to you personally?
    \item What motivates you to do this work on a regular basis?
\end{itemize}

\item \textbf{Experiences with running the events.}
\begin{itemize}
    \item How do you go about organizing a typical event you run in your organization?
    \item What are the most challenging aspects of this experience?
    \item Which outreach strategies do you use to promote your events?
    \item How many volunteers and supporting staff do you have to help you run the events?
    \item Could you tell me about the funding that you require to run the events?
\end{itemize}

\item \textbf{Teaching methods.}
\begin{itemize}
    \item How do you choose the topic for a typical event you run in your organization?
    \item After you have selected a topic, how do you go about preparing to cover it during the event?
\end{itemize}

\item \textbf{Success metrics.}
\begin{itemize}
    \item How would you define ‘success’ in the work that you do?
    \item What is the overarching goal that your events are trying to achieve?
    \item What makes you say to yourself “that was a good event”?
\end{itemize}

\end{itemize}

\subsection{Interviews with participants}
Additionally, we aim to evaluate the efficacy of these educational initiatives—including workshops, meetups, consultations, and other activities—organized by advocates in leading to the adoption of secure online behaviors by members of Black communities. To this end, we will conduct interviews with participants of digital safety initiatives to understand their online security and privacy attitudes, needs, and concerns, as well as their experiences with the advocate-led initiatives.  Both the interviews with community-level advocates and participants will help foster our understanding of how to address security- and privacy-related needs more effectively.

\subsubsection{Subject Population}
We will use the following criterion to select participants to interview:

\begin{enumerate}
\item A “participant” is someone who previously took part in an initiative organized by an “advocate.”
\end{enumerate}

We aim to interview 3 to 5 participants per single initiative, and we expect to focus on at least three different types of initiatives. We will ask Mitchell and Aoun, as well as other advocates we interview, to help us reach out to the attendees of their events by sending an invitation to our screening survey. As before, respondents will answer demographic questions and indicate their preferred contact information, allowing us to reach out to them directly to schedule the interview. At the end of the interview, we will compensate each interviewee with \$20 for their participation in the study.

\subsubsection{Interview Guide}
We will conduct semi-structured interviews lasting around 30 minutes to explore the experiences of participants attending advocate-led initiatives and their perceptions and attitudes of digital safety and privacy. In particular, we will focus on the following three topics:

\begin{itemize}

\item \textbf{Background and finding out about events.}
\begin{itemize}
    \item How did you find out about the workshop in the first place?
    \item What motivated you to attend the workshop?
    \item Was there anything specific that you wanted to get out of attending the workshop?
    \item Have you attended workshops from this organization or another one since the first workshop?
\end{itemize}

\item \textbf{Experiences with participating in the events.}
\begin{itemize}
    \item Could you tell me about your experiences attending one of the workshops?
    \item Based on your experience attending the workshop, what were the key lessons that you learned?
    \item How easy or hard was it for you to understand the material taught in the workshop?
    \item How helpful or unhelpful did you find the course material taught in the workshop?
    \item Would you recommend this workshop to a friend or a colleague?
\end{itemize}

\item \textbf{Privacy and security threats.}
\begin{itemize}
    \item What does ‘digital safety and privacy’ mean to you personally?
    \item When you think about ways you keep yourself safe, what things come to mind?
    \item What kind of threats or risks do you worry about?
    \item Were there any specific concerns about your digital safety or privacy that led you to attend the workshop?
    \item Are there any other concerns that you felt were not answered by attending these events?
\end{itemize}

\end{itemize}

\subsection{Ethics}
We will record audio from the interviews for transcription, coding, and analysis. The raw audio recordings will be securely stored and subsequently deleted as soon we finish transcribing them. We will also remove any sensitive and personally identifiable information contained in the interviews as part of the transcription process. Additionally, we will assign unique random identifiers to connect the survey responses and interview transcripts to the same participants; the survey responses, interview data, and the identifiers will be stored separately from any personally identifiable information. 

Our team has done extensive human subjects research and has years of research experience in the privacy and security domain. All members of our research team have performed or will perform the Responsible Research and Social \& Behavioral Research CITI Program training, or equivalent. The study is currently undergoing the Institutional Review Board (IRB) review process at the University of California, Berkeley (under the protocol ID: 2021-02-14070). 

\section*{Acknowledgments}

We acknowledge the financial support of the Center for Technology, Society \& Policy (CTSP) and the Center for Long-Term Cybersecurity (CLTC) at the University of California, Berkeley. Additionally, we would like to thank Matt Mitchell, Sarah Aoun, and Sarah Chasins for their assistance with this project.

\bibliographystyle{plain}
\bibliography{paper}

\end{document}